\documentclass[12pt,preprint]{aastex}
\begin{document}

\title{The NGC  672 and NGC 784 Galaxy Groups: Evidence for Galaxy Formation and Growth Along a Nearby Dark Matter Filament}

\author{Adi Zitrin\altaffilmark{} and Noah Brosch\altaffilmark{}}
\affil{The Wise Observatory and the Raymond and  Beverly Sackler School of Physics and
Astronomy, the Faculty of Exact
Sciences, \\ Tel Aviv University, Tel Aviv 69978, Israel}
\email{adiz@wise.tau.ac.il, noah@wise.tau.ac.il}

%\maketitle
%\onehalfspacing
%\begin{singlespace}
%\end{singlespace}
%\twocolumn
%\columnsep 10pc

\begin{abstract}
%\begin{footnotesize}
We present U, B, V, R, I, H$\alpha$ and NUV photometry of 14 galaxies in the very local Universe (within 10 Mpc). Most objects are dwarf irregular galaxies (dIrr) and are probably associated with the NGC 672/IC 1727 and NGC 784 galaxy groups. The galaxies are at low redshift (51$\leq$v$_{\odot} \leq$610 km sec$^{-1}$) and most appear projected on the sky as a six degree long linear filament. We show that the galaxy positions along this filament correlate with their radial velocity, hinting to an interpretation as a single kinematic entity. Our CCD photometry indicates that all objects qualify as ``dwarf galaxies'' with M$_B\geq-18$ mag. We examine the star formation (SF) properties of individual objects in the context of their immediate environment. The current SF rate (SFR) is derived directly from the H$\alpha$ line flux. An approximate SF history is derived by comparing the multi-band photometry with results of galaxy evolution models from Bruzual \& Charlot (2003a, 2003b), assuming short SF bursts separated by long quiescence periods.

Relations between the current SFR and the HI mass or the absolute B magnitude for the galaxies in these groups indicate that these objects behave like normal galaxies. A comparison of the photometric measurements with evolutionary synthesis model predictions indicates that most objects can be understood as containing at least one ``old'' stellar population ($\geq$1-10 Gyr) and one ``young'' population ($\leq$30 Myr). For both groups, the recent SF bursts appear to have occurred at similar times, a few to a few 10s of Myr ago, arguing for synchronicity in star formation in these objects.

In an attempt to evaluate the possible role of galaxy-galaxy interaction, we investigate the trend of the SFR with an object's projected distance from the brightest and most massive galaxies of each group. We do not find a steadily decreasing star formation as function of this distance; such a result could be expected if the star formation would have been triggered by interactions. We propose that one possible explanation of the $\sim$synchronous star formation in all objects is accretion of cold gas from intergalactic space onto dark matter haloes arranged along a filament threading the void where these dwarf galaxies reside. We point out this galaxy sample as an ideal target to study hierarchical clustering and galaxy formation among very nearby objects.

%It has been known that generally, mergers or interactions between normal galaxies tend to enhance their SF (see Larson \& Tinsley 1978, Li et al.\ 2008). It is also common to believe that at low redshifts, galaxies in groups show somewhat a reverse relation, manifested in a higher SFR the more isolated they are (see Martig \& Bournaud 2008). Our findings support and bridge these conclusions with some refinement. The strongly interacting normal galaxies show indeed enhanced SFR. The dIrr galaxies in the immediate environment of the larger galaxies of each group, show decreasing SFR and SFR surface density with the increase in distance from the main galaxies. Only further away, would more isolated normal galaxies show again an expected higher SFR.

%\end{footnotesize}
\end{abstract}


%{\bf Still to do: add pix of all galaxies (B+nH$\alpha$)}

\section{Introduction}
Star formation is one of the more important processes in the Universe. Studying both the SF history in galaxies and the influence of the environment on this process contributes to the understanding of cosmic evolution. In the past few decades increasing efforts have been made to better understand the dependence of SF on local and environmental properties. Tidal interactions were promoted as triggers and enhancers of SF (e.g., Larson \& Tinsley 1978, Li et al.\ 2008). In practice, the influence of tidal interactions is more complex, depends upon various factors, and varies among different environments. Hashimoto et al.\ (1998) studied the variation and dependence of these influences in different environments for different types of galaxies and found different SFRs in the field and in clusters, even when the galaxy densities are similar. They also found that for a given galaxy concentration index, galaxies in lower-density environments show higher SFR levels than objects in higher galaxy density regions. At least two processes were found to influence the susceptibility of SF to the environment: gas-removal processes responsible for the variation with galaxy density of the SFRs of normal galaxies, and galaxy-galaxy interactions responsible for the prevalence of SF bursts in intermediate-density environments. At low redshifts, galaxies in groups may show a reverse relation with a somewhat higher SFR (but not extreme burst conditions) the more isolated a galaxy is (see Martig \& Bournaud 2008). However, Brosch et al.\ (2004) examined specifically the influence of interactions on SF in dwarf galaxies and concluded that galaxy interactions are probably not a primary trigger of present SF in such objects.

About half the galaxies in the Universe are probably in groups (Mamon 2007) and these constitute an interesting environment for the SF investigation. The evolution of galaxies within groups may depend upon the density of the group and the level of interaction between its galaxies. Many groups consist of normal galaxies surrounded by clouds of dwarf galaxies (Miller 1996, Cote et al.\ 1997, Mateo 1998). The dwarf galaxies (DGs) may have formed due to tidal interaction of individual galaxies or of groups of galaxies, or as leftover material from the colliding and merging of more massive galaxies (e.g., Hunsberger et al.\ 1996, Bournaud \& Duc 2006). They may also represent early stages of hierarchical clustering, with the stars that provide the observed luminosity forming in a low-mass halo.
 %models supported by supernova winds and dark matter considerations (e.g., Dekel \& Silk 1986, Nagashami \& Yoshii 2004).

The DGs in a group contain usually only a few percents of the group mass and in many cases may be accreted by larger galaxies. Although dwarf galaxies constitute the majority of the galaxy population, large uncertainties surround their formation and evolutionary histories. The dwarf galaxy population follows a strong morphology-density relation, with passively evolving systems mostly found in close proximity to massive galaxies, in contrast to the more widespread gas-rich,
star forming population. Studying galaxy groups, in particular the  dwarf-rich ones, contributes to the understanding of cosmic structure and evolution, since such groups constrain cosmological models, imply the shape of the dark halos around the massive galaxies, and play a role in the evolution of these galaxies (see Bournaud \& Duc 2006 and references therein).

The present study complements similar ones on other groups of galaxies in the local Universe, such as for the NGC 628 and M81 groups (e.g., Sharina et al.\ 2006, Karachentsev \& Kaisin 2007). Hickson (1982) published a catalog of compact groups of galaxies in the local Universe, and investigated various characteristics of these groups (e.g., Hickson et al.\ 1992, de Oliveira \& Hickson 1991, 1994). Samples of galaxies in other environments have been examined, such as in the Virgo cluster (e.g., on dwarf galaxies, Almoznino \& Brosch 1998a, b; Brosch, Heller \& Almoznino 1998; Heller, Almoznino \& Brosch 1999; Heller \& Brosch 2001). The influences of galaxy interactions on star formation were reviewed by Larson \& Tinsley (1978), Hashimoto et al.\ (1998) and Li et al.\ (2008). In most previous papers the galaxies were either normal (i.e., large spirals, lenticulars or ellipticals) or dwarf galaxies; the specific aspect the present paper adopts is to study these properties in a fairly isolated group or groups of DGs.

Tidal interactions between two strongly-interacting major galaxies tend to form sometimes rather concentrated clouds of DG satellites around them, or long tails of DGs as seen e.g., in the Hickson compact group 100 (de Mello et al. 2008). If one would observe dwarf galaxies in a group apparently aligned on the sky in a linear configuration, and with no major galaxies in the immediate neighborhood, this could imply the presence of a dark matter filament onto which the intergalactic matter collapses, forming the observed DGs.

This work studies a sample galaxies in the very local Universe ($\sim$10 Mpc)  consisting of the NGC 672/IC 1727 group of galaxies, the NGC 784 group of galaxies, and several other galaxies in the same vicinity with similar redshifts. Our attention was drawn to this sample when inspecting the precursor observations of the ALFALFA survey (Giovanelli et al. 2005) because of its obvious linear structure that could qualify as a ``filament'' of galaxies, and because of its apparent isolation. The galaxies are located approximately in the anti-Virgo direction, in a region of the Universe that is a void in the galaxy distribution; this is the ``Local Mini-void'' (Karachentsev et al. 2004). As will be shown below, all galaxies are ``dwarfs'' (M$_B\geq$-18 mag) and most are dIrr (e.g., Karachentseva \& Karachentsev 1998, Huchtmeier et al. 2000). We study the SF properties and SF history of each galaxy using surface photometry, and examine general similarities and dependencies on the local, environmental and mutual properties.

Current $\Lambda$CDM simulations predict that low-amplitude filamentary structures criss-cross the voids (Peebles 2007). The galaxies corresponding to the halos in those filaments are expected to be low-luminosity, star-forming galaxies (Hoyle et al. 2005). Saintonge et al. (2008) analyzed the ALFALFA catalog covering a portion of the nearby void in front of the Pisces-Perseus Supercluster at cz$\sim$2000 km sec$^{-1}$ and, within a volume of 460 Mpc$^3$, did not detect a single galaxy. In contrast, one could expect to detect 38 HI sources in such a volume based on scaling the predictions of Gottl\"{o}ber et al. (2003) with a dark-to-HI mass ratio of 10:1.

Dekel \& Birnboim (2006) discussed the bimodality observed in galaxy properties about a characteristic stellar mass of $3\times10^{10}$ M$_{\odot}$. The bimodality implies that less massive galaxies tend to be ungrouped, blue, star-forming disks and would be formed in low galaxy-density regions, while more massive galaxies are typically grouped, red, old-star spheroids and would reside in clusters and in denser groups. In haloes below a critical shock-heating mass of 10$^{12}$ M$_{\odot}$ disks are built by cold-gas streams, not heated by a virial shock, yielding efficient and prompt star formation. The Dekel \& Birnboim scenario explains naturally why one could expect to see dwarf galaxies even in regions that are not conducive to intense tidal interactions, provided dark matter haloes of the right mass are present and HI material for SF is available.

In what follows we argue that the objects studied here may be the first case in the local Universe where we witness the first stages of aggregation of matter that will, eventually, form a major galaxy. The plan of the paper is as follows: in Section~\ref{sec.sample} we describe the galaxy sample, the observations, mostly collected at the Wise Observatory, and the data reduction, in \S~\ref{sec.results} we present the results and discuss individual objects, in \S~\ref{sec.discuss} we discuss the results and present arguments in favour of the interpretation proposed above, and in \S~\ref{sec.summary} we conclude and summarize our findings.

\section {Observations and data reduction}
\label{sec.sample}

\subsection {The sample}

The sample consists of the catalogued NGC 672/IC 1727 and NGC 784 galaxy groups and of several additional galaxies in the same sky area and at similar redshifts; these are listed in Table \ref{gal}. The galaxies were chosen primarily from the ALFALFA survey (see Giovanelli et al.\ 2005, Saintonge et al.\ 2008). %This 21cm line survey is in its third year of mapping 7074 square %degrees of the high galactic latitude sky, and it is being performed with the Arecibo radio %telescope, with the ALFA focal plane array. Most of the galaxies were chosen from the %precursor ALFALFA survey results (see Giovanelli et al.\ 2005). That survey
ALFALFA is an unbiased HI survey of the extragalactic sky visible from Arecibo. The precursor ALFALFA observations (Giovanelli et al.\ 2005) covered the sky area  (0h$\leq$RA$\leq$6h) and (26$^{\circ} \leq \delta\leq 28.5^{\circ}$) and 10 galaxies in the vicinity of NGC 672/IC 1727 with similar redshifts (cz$\leqslant$600 km sec$^{-1}$) were included among the 166 identified objects.

\begin{figure}[t]
\centering{
  \includegraphics[width=12cm]{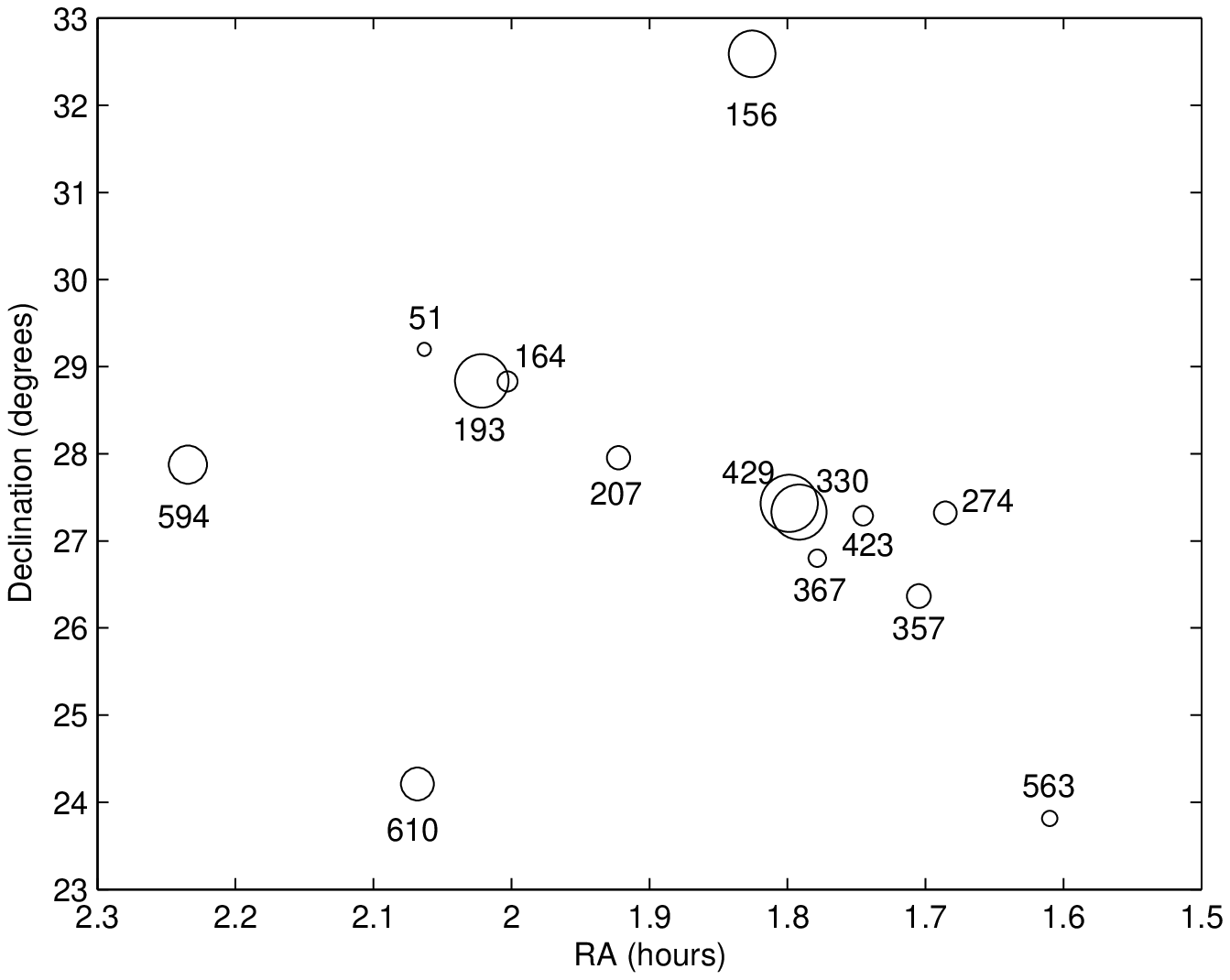}
  \caption{The galaxies as projected on the sky, represented by circles of different sizes. The circle sizes are proportional to the galaxies' semi-major axes. Redshifts are marked next to each galaxy symbol. A linear structure is evident from the distribution of 11 of the 14 objects plotted here.}
  \label{fig:galaxies}}
\end{figure}

Figure \ref{fig:galaxies} shows the projection of the different objects on the sky, with the redshift of each object indicated. It is clear that the galaxies form some kind of linear structure, which we call ``filament''; the projected distribution is approximately linear with the objects at the north-east side showing generally lower redshifts. The length of the filament is $\sim$six degrees; a few galaxies diverge significantly from the linear distribution and we consider them to be ``isolated''. The distance to the galaxies is of order 5 Mpc; the $\sim6^{\circ}$ projected angular extent translates into a projected physical length for the linear structure of $\sim$500 kpc. %{\bf Here it would help to have a plot of the position of a galaxy along a best-fitted line to the RA,DEC of the objects, against the redshift}

As already mentioned, most of the galaxies studied here were previously classified as dwarf irregulars (e.g., Karachentseva \&  Karachentsev 1998, Huchtmeier et al.\ 2000). In Table~\ref{gal} we list not only the galaxy name, which in some cases is taken from the Arecibo Galaxy Catalogue (AGC: a private compilation of Haynes and Giovanelli maintained at Cornell), but also the heliocentric velocity in km sec$^{-1}$ and the 50\% width of the HI line (also in km sec$^{-1}$) taken from Saintonge et al. (2008) or from the HyperLEDA data base (Paturel et al. 2003). Column 7 lists the position angle (PA) of the major axes of the objects, determined from a visual inspection of the blue galaxy images on-line at the Canadian Astronomy Data Centre (CADC). The PA is measured in degrees, clockwise from North through West. AGC  111945 was too faint and small to derive its PA. The last column of the table lists the galaxy group to which an object could belong.

\begin{table}[!h]

\caption{The sample galaxies arranged by group membership}
\vspace{0.5cm}
\label{gal}
\begin{footnotesize}
\begin{center}
\begin{tabular}{|c|c|c|c|c|c|c|c|}
\hline
Galaxy & ALFALFA name & $\alpha$[hh:mm:ss.s] & $\delta$[$^\circ$:':''] & v$_{\odot}$ & w$_{50}$ & PA ($^\circ$) & Group\\
\hline
NGC 672 & HI014753.9+272555 & 01:47:54.5 & +27:25:58 & 429 & 205 & 332 & NGC 672\\
IC 1727 & HI014729.9+271958 & 01:47:29.9 & +27:20:00 & 330 & 115 & 35 & NGC 672\\
AGC 110482 & HI014214.9+262202 & 01:42:17.3 & +26:22:00 & 357 & 30 & 53 & NGC 672\\
AGC 111945 & HI014441.4+271707 & 01:44:42.7 & +27:17:18  & 423 & 36 & ? & NGC 672\\
NGC 111946 & HI014640.9+264754 & 01:46:42.2 & +26:48:05 & 367 & 21 & 0 & NGC 672\\
AGC 112521 & HI014105.8+272007 & 01:41:08.0 & +27:19:20 & 274 & 26 & 291 & NGC 672\\
LEDA169957 & --- & 01:36:35.9 & +23:48:54 & 563 & 47 & 307 & Isolated\\
NGC 784 & HI020115.0+284953 & 02:01:16.9 & +28:50:14 & 193 & 88 & 0 & NGC 784\\
AGC 111977 & HI015519.2+275645 & 01:55:20.4 & +27:57:13 & 207 & 29 & 297 & NGC 784\\
AGC 111164 & HI020009.3+284954 & 02:00:10.2 & +28:49:53  & 164 & 27 & 24 & NGC 784\\
UGC 1281 & --- & 01:49:32.0 & +32:35:23 & 156 & 98 & 306 & NGC 784\\
AGC 122834 & HI020347.0+291153 & 02:03:47.0 & +29:11:53 & 51 & 10 & 297 & NGC 784?\\
AGC 122835 & HI020533.0+291358 & 02:05:33.0 & +29:13:58 & 50 & 23 & - & N784; no Opt ID \\
UGC 1561 & --- &02:04:05.1 & +24:12:30 & 610 & 47 & 90 & Isolated\\
NGC 855 & HI021404.3+275302 & 02:14:03.6  & +27:52:36 & 594 & 81 & 305 & Isolated\\
\hline
\end{tabular}
\end{center}
\end{footnotesize}
\end{table}

Though we found no previous results covering the SF in the sample galaxies, some objects were investigated before: the NGC 672/IC 1727 pair and their interaction have been studied by de Vaucouleurs et al.\ (1976), Combes et al.\ (1980), Hodge \& Kennicutt (1983), Sohn \& Davidge (1996), Karachentseva \& Karachentsev (1998);  NGC 784 and some members of its group have been studied by Karachentseva \& Karachentsev (1998), Huchtmeier et al.\ (2000), Tikhonov \& Karachentsev (2006), Tully et al.\ (2006); other galaxies of the sample were studied by e.g., Wallington et al.\ (1988), Huchtmeier \& Richter (1989), Moustakas \& Kennicutt (2006). Most galaxies were measured in HI by the ALFALFA survey (e.g., Giovanelli et al.\ 2005) or are included in Karachentsev's catalog of neighboring galaxies (Karachentsev et al. 2004). Saintonge et al.\ (2008) studied the HI properties of the NGC 672/IC 1727 and NGC 784 galaxy groups among other objects.

Saintonge et al.\ (2008) included the NGC 672/IC 1727 and the NGC 784 groups along with two new candidate members of the NGC  784 group among the objects they studied. Only one of those two candidates, AGC 122834, has an optical counterpart. The other, AGC 122835 at 02:05:33.0 +29:13:58, is also listed in Table~\ref{gal}. In addition to the 12 objects retrieved from the ALFALFA data sets, NED was queried for objects up to 300 arcmin away and with similar velocities; this added three additional neighboring galaxies. All the objects fall within the ``Local Mini-void'' Karachentsev et al.\ (2004, Fig. 4). Note that the ALFALFA survey is likely to find even more members of this filament once more survey data for other declination strips will be released and redshifts of low surface brightness objects will be measured.
%It is reasonable to assume that galaxies which are close in projected distance may actually be at the same distance. Such an assumption may be justified by the sky distribution of the sample galaxies; these appear to form a linear distribution.
%The redshifts are commensurate with the velocity dispersion of field galaxies, thus assuming that distances are simply related to redshifts may mask possible relationships.

The literature contains a number of distance estimates to objects in our sample. Sohn \& Davidge (1996) assigned a distance of 7.9$^{+1.0}_{-0.9}$ Mpc to NGC 672 based on the brightness of the red supergiants. Karachentsev et al.\ (2004) reported 7.2 Mpc to the NGC 672/IC 1727 pair based on the luminosity of the brightest stars. A distance of 5.0 Mpc was measured in a similar way for NGC 784 by Drozdovsky \& Karachentsev (2000). The distance to AGC 111977 and AGC 111164 was reported as 4.7 Mpc by Karachentsev et al.\ (2004) using the red giant branch method. The distances to UGC 1281 was estimated by Karachentsev et al.\ (2004) to be 5.4 Mpc using the brightest stars. The reported distance for NGC 855 (using the surface brightness fluctuations method) was 9.73$\pm$1.7 Mpc (Tonry et al. 2001).

The question of distances to these galaxies was discussed by Giovanelli et al. (2005) who used the peculiar velocity model of the local Universe of Tonry et al. (2000). The model provides the expected peculiar velocity at a given point in space, but can be inverted to yield the distance to a galaxy at certain celestial coordinates and with a measured redshift. Giovanelli  al. caution that the ``thermal'' rms peculiar velocity of galaxies in the Tonry et al. model translates into a distance uncertainty of about 2.7 Mpc and thus the distances to nearby objects, and in particular to the NGC 672 and NGC 784 groups, are highly uncertain.

Since the distance estimates were produced by a variety of methods, one wonders whether the distance differences between members of the sample is believable. In particular, when Karachentsev et al.\ (2004) quote 4.7 Mpc to AGC 111977 and AGC 111164 and 9.73 Mpc to NGC  855, can we indeed assume that these objects are separated by $\sim$5 Mpc in depth? Is it possible that the objects could be at about the same distance, despite their somewhat different redshifts?

Given the uncertainty in distance, we decided to disregard peculiar velocities and use distances assuming only Hubble expansion with H$_{0}$=73 km sec$^{-1}$/Mpc to evaluate intrinsic parameters for each galaxy. The HI masses and the SFRs were calculated assuming these distances. However, in the discussion of the 3D structure traced by the galaxies we retain the option of considering the objects to be aligned along a filament (that may be inclined to the line of sight) and being at approximately the same distance from us.

%{\bf This only represents a small change to the velocities, since all the group galaxies are %almost in the same place, this is where the 0.6 Mpc comes from. I propose deleting this entire %paragraph.} We are aware that distances to objects at such small redshifts cannot be %determined by assuming simple Hubble expansion. However, distances calculated by peculiar %velocity field (as given in Saintonge et al.\ 2008 and references therein) show similar %consistency, with distances larger by about $\sim$0.6 Mpc than those assuming Hubble %expansion.  Moreover, in order to examine the reliability of our results, they were %recalculated with the two optional distance measurements (by velocity field published in %Saintonge et al.\ and Karachentsev's distances). This maintained the general consistency of %the results and the figures shown throughout this paper.

\begin{figure}[t]
\centering{
  \includegraphics[width=12cm]{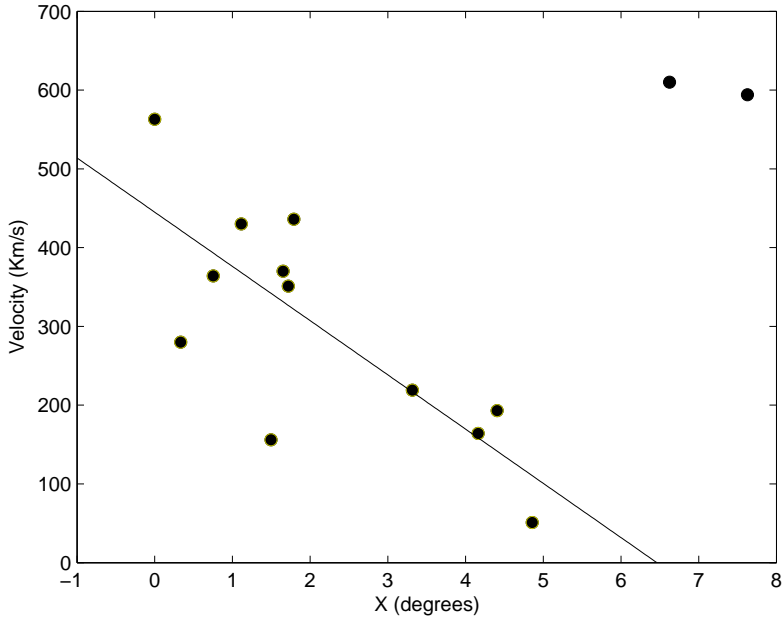}
  \caption{Position-velocity diagram for all the galaxies as projected on the sky. The horizontal axis marked X is the distance along a line fitted to the galaxy positions. The two symbols at the top-right part of the diagram are UGC 1561 and NGC 855.}
  \label{fig:gals_X_vs_Vel}}
\end{figure}

We plot in Figure~\ref{fig:gals_X_vs_Vel} the position of each galaxy along the line fitted to their projected location on the sky vs. the heliocentric radial velocity of each object. The plot shows that most galaxies line up nicely on a $\sim$6$^{\circ}$ linear feature, with those at higher Right Ascension having generally a lower radial velocity. This kinematic relation can be understood as additional evidence that the objects are probably part of the same large structure in the nearby Universe. The two points at the upper right in Figure~\ref{fig:gals_X_vs_Vel} are UGC 1561 and NGC 855; it is possible that these might not be part of the same kinematic structure delineated by the other galaxies.

\subsection {Observations}

All broad-band and H$\alpha$ observations were performed with the Wise Observatory (WiseObs) 1-m telescope between September 2006 and February 2008. We obtained CCD images of all 14 galaxies of the sample using the PI VersArray camera with a $1340\times1300$ pixel thinned and back-illuminated CCD and a scale of 0.58"/pixel at the f/7 Ritchey-Chr\'{e}tien focus. The CCD saturation level is 65535 counts per pixel, the gain is 2.1 $e^{-}$/ADU, and the readout noise is 2.87 $e^{-}$ per pixel. The seeing on most nights was 2.2"-2.8". Images of each galaxy were obtained through UBVRI and zero-redshift H$\alpha$ filters. Landolt (1992) and spectrophotometric (e.g., Oke 1990) standard stars were imaged along with the galaxy fields on nine photometric nights for H$\alpha$ and broad-band calibration. Most of the galaxies were calibrated on more than one night and yielded a typical photometric error of 0.06 mag for all broad bands.

Typical exposure times were 20 minutes for the U and H$\alpha$ filters, and 10 minutes for the other broad-band filters. Each galaxy was imaged at least three times through each filter, but for most galaxies 4-5 images per filter were obtained. Thus, at least 30 minutes of integrated exposure time were collected through the B, V, R and I filters, and at least 60 minutes through the U and H$\alpha$ filters.

The WiseObs set of filters was replaced between the two sessions - fall of 2006 and fall of 2007. Due to instrumental problems we discarded all U-band images taken in 2006. Three galaxies were imaged and calibrated through the new U-band filter in 2007/8. Note also that all long-exposure images in the I band showed interference fringes. These were removed by an interactive script that uses the IRAF \textit{rmfringe} command and an image of the fringe pattern with no objects in it, created by median-combining many night-time I-band images of different sky fields.

The H$\alpha$ filter used has $\lambda_{c}=6559\pm$2{\AA}, FWHM=56$\pm$2{\AA}, peak transmittance 58.9$\pm0.3\%$, and out-of-band transmittance $\leq$0.06\% (Spector 2006).
Note that the measured H$\alpha$ line flux includes, in principle, the [NII] lines at 6548{\AA} and 6584{\AA}. Since the ratio of [NII] to H$\alpha$ line fluxes in dwarf irregular galaxies is $\leq$10\% (e.g, Kennicutt 1983), we disregard the possible [NII] contribution through the $H\alpha$ filter. This is specifically justified for the four objects in our sample with global spectrophotometry from Moustakas \& Kennicutt (2006), these all have [NII]/H$\alpha \leq$0.08.

The galaxies of the sample have radial velocities $\leq$610 km sec$^{-1}$; this translates into a minor redward shift of the H$\alpha$ emission line ($\leq12${\AA}), for an effective filter transmission of at least $\sim95\%$ of the peak transmittance. Since the galaxies are at $\sim$zero redshift, the R filter was used for continuum measurement. Assuming a pessimistic photometric error of 0.1 mag, and adding in quadrature the possible [NII] contribution and the filter transmission contribution, the resulting H$\alpha$ flux measurement accuracy is 0.15 mag. This increases to 0.18 mag for the net-H$\alpha$ calculation (see below) due to the photometric error in the R band. For some galaxies, where both R
and H$\alpha$ have low photometric errors, the real calculated photometric error is smaller than this ``typical'' value.

We collected redshifts and HI line fluxes, which were converted to mass using $M_{HI} = 2.356 \times 10^{5}D^{2}F_{c}$ with $M_{HI}$ in solar units and $D$ in Mpc, from the ALFALFA survey (see Giovanelli et al.\ 2005, Saintonge et al.\ 2008) and other published data (e.g, Huchtmeier \& Richter 1989, Karachentsev et al.\ 2004, NED). NUV images for most of the sample galaxies were retrieved from the GALEX archive (\textit{http://galex.stsci.edu/GR4/}). The galaxies were measured with the same aperture as the broad-band images, and the measurements were calibrated using the formal GALEX procedure (Morrissey et al.\ 2007). The GALEX NUV band has an effective wavelength of 2315.7{\AA}, a peak at 2200{\AA}, and effective bandwidth is from 1771{\AA} to 2831{\AA} (Morrissey et al.\ 2007). We assume similarity between this filter and the 2200{\AA} filter calculated in the Bruzual \& Charlot models (2003b) and use the derived NUV magnitudes to construct a (NUV-V) colour index. The typical photometric error of this colour index is similar to those of the other indices used here.

\subsection {Galaxy photometry}
The galaxies were measured using the IRAF ``Ellipse'' function at a surface brightness level of $\mu_B \sim25.5$ mag arcsec$^{-2}$. The sky background level was determined using a galaxy-dependent method. For each image, the sky background was measured in an elliptical annulus whose inner semi-major axis (SMA) was $\sim$1.35 times the primary SMA and its outer SMA was $\sim$1.75 times the primary SMA. The ``Primary SMA'' term refers to the SMA of the elliptical aperture within which the galaxy was measured (at a brightness level of $\mu_B \sim25.5$ mag arcsec$^{-2}$).

The final result for each band was corrected for extinction and for a possible colour term. The results were also corrected for Galactic extinction (GE) according to the prescriptions of Schlegel et al.\ (1998). Since the galaxies are in a similar sky area, the typical Milky Way (MW) extinction values for this area are $A_{U}\sim0.4$, $A_{B}\sim0.34$, $A_{V}\sim0.26$, $A_{R}\sim A_{H_{\alpha}}\sim0.2$, and $A_{I}\sim0.17$, but exact values were calculated  specifically for each galaxy. For the GALEX NUV measurement the typical MW extinction was $\sim0.9$ mag. The error in the determination of the extinction values is $\sim10\%$.

The continuum flux contribution through the R filter was subtracted from the H$\alpha$ measurement to provide the net-H$\alpha$ (nH$\alpha$) line flux. The derived nH$\alpha$ was subtracted from the average $R$ flux, thus the final $R$-band magnitudes listed below are free of H$\alpha$ line emission. The nH$\alpha$ images were produced by subtracting the R image from the H$\alpha$ image, after normalization by the averaged total flux of the reference stars. These nH$\alpha$ images were used to find, count and measure individual HII regions.

The photometry results, H$\alpha$ fluxes, derived SFR (see below) and HI mass data are presented in Table~\ref{T3}. We compared our derived photometry with published values and found reasonable compatibility. Four of our objects were observed by Moustakas \& Kennicutt (2006) in their scanning-slit spectrophotometric survey of spiral galaxies. For these objects we found also reasonable correspondence with the Moustakas \& Kennicutt values for the H$\alpha$ flux being, on average, 8\% brighter than ours.

Note that the brightest objects of the sample, NGC 672 and NGC 784, have absolute magnitudes -16.9 and -16.4 if located at a distance of 5 Mpc, and could be considered ``dwarf galaxies'' even if at twice the distance. This demonstrates that the sample is indeed composed exclusively of low-luminosity galaxies.

\subsection {Comparison to galaxy evolution models}
The SFH of each galaxy was derived by comparing the observed global colours to predictions of galaxy evolution models  (GEMs). The GEMs are based on the GALAXEV library (Bruzual \& Charlot 2003a, 2003b). Since the main SF characteristic in dwarf galaxies is that it is generally proceeding in bursts, we selected to compare our observational results to model predictions of stellar populations formed in bursts. We used the 26 default models pre-calculated in GALAXEV using the Padova 1994 evolutionary tracks: 13 models use the Salpeter (1955) IMF and 13 the Chabrier (2003) IMF. No significant differences were found between the results obtained with either of the two IMFs. The comparisons were done for the optical colours [(U-B), (B-V), (V-R) and (V-I)], as well as for the (H$\alpha$-V) colour, which is produced in a procedure described below, and for the predicted (2200{\AA}-V) colour assumed to be identical to the (GALEX NUV-V) colour. %This %colour was %used for model comparison as well.

Using the H$\alpha$-V colour enhances the fit reliability since the line emission is determined by the amount of young massive stars in a galaxy. The models predict the number of LyC photons and we used this and the V magnitude to derive a distance-independent colour for each galaxy. The H$\alpha$ luminosity is related to the LyC photon flux (Osterbrock 1989):
\begin{equation}
N_{c} = 7.43 \times 10^{11}L(H\alpha),
\end{equation}
where $L(H\alpha)$ is in erg/s, which leads to:
\begin{equation}
H\alpha-V = 129.8 -2.5 \times log[N_{c}] - V_{abs},
\end{equation}
where both $log[N_{c}]$ and $V_{abs}$ [mag] are calculated in the model.

All models use standard lower and upper mass cutoffs of $m_{L}=0.1M_{\odot}$ and $m_{U}=100M_{\odot}$, and differ in metallicity, which ranges from 0.0001 to 0.05. The models calculate spectra and galactic colours at 220 time steps, from t=0 to t=20 Gyr. They were used to create a colour data base for each time step and for each metallicity. Since the models describe a single generation of stars formed in an instantaneous SF burst that only ages after being formed, a script was written to find the best $\chi^{2}$ fit of any combination of colours and burst times from the data base. The minimal $\chi^{2}$ criterion calculated by the script is given by:
\begin{equation}
{\rm Min}(\chi^{2}) = {\rm min} [\sum (\Delta colour)^{2}/err^{2}] / (N_{colours} - N_{bursts}).
\end{equation}
where $\Delta colour$ is the difference between each measured colour and the linearly-combined model-produced colour, $err$ is the error of the measured colour, and the sum refers to the measured colours (NUV-V, U-B, B-V, V-R, V-I and H$\alpha$-V). $N_{colours}$ is the number of available colours and $N_{bursts}$ is the number of bursts in the allowed combination.

Here we used primarily a combination of two bursts and the variables are the percentage of light produced by each population, the metallicity of this population, and the time of the burst. Since we fit essentially five predicted variables (fraction of light from the first burst in the observed brightness, the times of each of the two bursts, and the metallicity of each burst) to five measured colours [(U-B), (B-V), (V-R), (R-I) and (H$\alpha$-V)], the fit should be, in principle, fully determined. For cases where we have also (NUV-V), the fit to two bursts should be overdetermined. Instances where we lack U-band photometry cause the fit to be under-determined.

For some cases we were forced to assume more than two SF bursts; it is probable that these fit results are not unique. The script outputs not only the best fit, but also the next nine fits to determine the spread of the multiple solutions. The dispersion of the nine additional results around the best fit, in terms of the burst times, burst weight, and metallicities, was used as the error estimate.

The dispersion around the best fit was typically 1-3 Myr for the very recent bursts, whereas for the earlier bursts that occurred a few Gyr ago it was $\lesssim$1 Gyr. For very ``old'' bursts, $\sim$10 Gyr, the dispersion is high and the time resolution is low; we therefore indicate that these SF bursts took place at least 10 Gyr ago without specifying exactly when. In addition, there are sometimes two different possibilities described by the 10 best fits, which have similar $\chi^{2}$ values and we list both options below. The dispersion of the relative weights of the bursts (fraction of light measured now) is $\sim10\%$ around the best fit values.

The least constrained variable in the fitting procedure is the metallicity, since the observed colours can be constructed from bursts at two specific times, but with different metallicities. For a wide range of observed colours, different metallicity models shows differences smaller than the observational error. For this reason, the best fit sometimes indicates a galaxy metallicity decreasing with time and such apparently non-physical solutions are retained, since they could hint at an influx of extragalactic low-metallicity gas.

\section{Results}
\label{sec.results}

In this section we first present the results in terms of the integrated B-band photometry and colours listed in Table~\ref{T3}, then describe each object in a separate subsection.
Table \ref{T3} lists for each galaxy the name, B magnitude, (NUV-V), (U-B), (B-V), (V-R), (V-I), (H$\alpha$-V), H$\alpha$ line flux [erg~cm$^{-2}$~s$^{-1}$], distance [Mpc, assuming Hubble expansion with $H_{0}=73$\ km sec$^{-1}$/Mpc], HI mass as $logM_{HI}$ [M$_{\odot}$], and the SFR  [$M_{\odot}/yr$] as obtained from $SFR[M_{\odot}/yr] = 1.27\cdot10^{9}F_{H\alpha}\cdot D^{2}$, where $D$ is in Mpc, $F_{H\alpha}$ is in erg~cm$^{-2}$~s$^{-1}$ (Almoznino \& Brosch 1998b), and the numbers in square parentheses indicate the power-of-ten of the H$\alpha$ flux or of the SFR.%, and group membership association.
%{\bf Adi: where does the SFR formula come from (reference)?}

The best fits to the models and the derived SFH are presented in Table~\ref{T2}, listing the galaxy name, the time from the first burst in Gyr from the best model fit, and the dispersion of the nine following fits. If the fits suggest other options, these are listed as well, separated by a comma. The table then lists the fractional contribution of the second burst to the luminosity measured today, the metallicity of the second stellar generation, and the $\chi^{2}$ of the best fit. The models use the Padova 1994 track notations: m22, m32, m42, m52, m62 and m72 for metallicities of Z=0.0001, 0.0008, 0.004, 0.008, 0.02 and 0.05, respectively. For more information on notation see Bruzual \& Charlot (2003b).

%\begin{sidewaystable}[p]
%\begin{minipage}{160mm}
{\tiny
\begin{table}
\begin{center}
\caption{Photometry results [mag]. All broad-band magnitude errors were $\thicksim0.1$ mag or lower. The error in $logM_{HI}$ is typically 0.1}
\label{T3}
\begin{footnotesize}
\begin{tabular}{|c|c|c|c|c|c|c|c|c|c|c|}
\hline
Galaxy & Bmag & NUV-V & U-B & B-V & V-R & V-I & {H$\alpha$-V} & $F_{H\alpha}$ & $logM_{HI}$ & SFR  \\
\hline
NGC 672 & 11.55 & -0.23 & --- & 0.63 & 0.57 & 1.01 & 19.08 & $9.88\pm0.62)$[-13] &  9.04 & 3.65[-1]  \\
IC 1727 & 12.42 & -0.73 & --- & 0.50 & 0.46 & 0.74 & 18.76 & $5.34\pm0.85)$[-13] &  8.76 & 1.65[-2]  \\
AGC 110482 & 16.62 & -0.19 & --- & 0.58 & 0.48 & 1.04 & 17.48 & $3.88\pm0.38)$[-14] & 6.86 & 1.23[-3]  \\
AGC 111945 & 18.08 & -0.86 & --- & 0.62 & 0.66 & 0.40 & 18.85 & $2.98\pm0.93)$[-15] &  7.35 & 1.31[-4] \\
NGC 111946 & 18.08 & -0.54 & --- & 0.30 & 0.20 & 0.18 & 18.41 & $3.33\pm0.5)$[-15] &  6.64 & 1.09[-4]  \\
AGC 112521 & 17.88 & --- & 0.08 & 0.29 & 0.37 & 1.36 & 18.33 & $4.39\pm0.75)$[-15] & 6.35 & 8.20[-5]  \\
LEDA169957 & 17.87 & -0.54 & --- & 0.26 & 0.43 & 0.51 & 19.99 & $9.0\pm6.0)$[-16] &  7.4 & 6.80[-5]  \\
NGC 784 & 12.10 & -0.7 & --- & 0.01 & 1.05 & 0.9 & 17.05 & $2.2\pm0.22)$[-12]  &  7.85 & 1.60[-2]  \\
AGC 111977 & 16.71 & -0.61 & --- & 0.52 & 0.51 & 0.52 & 20.71 & $1.74^{+2.2}_{-1.74}$[-15] & 6.21 & 1.99[-5]  \\
AGC 111164 & 17.97 & --- & --- & 0.68 & 0.43 & 0.74 & 17.79 & $9.27\pm0.12$[-15] &  5.81 & 5.94[-5]  \\
UGC 1281 & 13.16 & --- & -0.22 & 0.67 & 0.40 & 0.73 & 19.14 & $2.23\pm0.76$[-13] &  7.71 & 1.63[-3]  \\
AGC 122834 & 19.81 & --- & 1.11 & 1.33 & 0.61 & 1.32 & 19.52 & $6.34\pm3.49$[-16] & 4.99 & 3.93[-7]  \\
UGC 1561 & 14.79 & -0.29 & --- & 0.18 & 0.51 & 0.75 & 17.23 & $2.59\pm0.23$[-13] &  7.72 & 2.30[-2]  \\
NGC 855 & 13.19 & 0.64 & --- & 0.7 & 0.55 & 0.95 & 18.22 & $5.16\pm0.91$[-13] &  7.78 & 4.34[-2]  \\
\hline
\end{tabular}
%\end{minipage}
\end{footnotesize}
\end{center}
%\end{sidewaystable}
\end{table}
}

%{\bf Note that the H$\alpha$ entries for some galaxies in Table~\ref{T3} are obviously wrong, %since the error should have been larger than listed}

\begin{table}[!h]
%\begin{minipage}{160mm}
\begin{center}
\caption{Model fitting results}
\label{T2}
%\vspace{1.5cm}
\begin{footnotesize}
%\hspace{-2.1cm}
\begin{tabular}{|c|c|c|c|c|c|c|c|}
\hline
Galaxy & 1$^{st}$ burst (Gyr) & Weight & Metal & 2$^{nd}$ burst (Myr) & Weight & Metal & $\chi^{2}$ \\
\hline
NGC 672 & $>10$ & 0.57 & m32 & $6.6\pm1$ & 0.43 & m72 & 2.67  \\
IC 1727 & $>10$ & 0.37 & m42 & $8.5\pm1$ & 0.63 & m72 & 1.79 \\
AGC 110482 & $>10$ & 0.3 & m72 & $6.3\pm1$ & 0.7 & m72 & 1.65  \\
AGC 111945 & $1.26\pm0.3$, $\sim6.3$ & 0.5 & m62 & $3.9\pm{3}$ & 0.5 & m22 & 9.17 \\
NGC 111946 & $1.5\pm0.3$, $\sim10$ & 0.5 & m52 & $2.5^{+3}_{-2.5}$ & 0.5 & m72 & 1.21  \\
AGC 112521 & $>10$ & 0.37 & m42 & $6.6\pm3$ & 0.63 & m62 & 1.89  \\
LEDA169957 & $>10$ & 0.35 & m32 & $20\pm5$ & 0.65 & m42 & 1.2 \\
NGC 784 & $>10$, $\sim2$ & 0.31 & m42 & $8.7\pm1$ & 0.69 & m32 & 16.78  \\
AGC 111977 & $>10$, $1.26\pm0.3$ & 0.65 & m42 & $27.5\pm{3}$ & 0.35 & m42 & 3.30  \\
AGC 111164 & $>10$ & 0.65  & m72 & $1.4\pm1$ & 0.35 & m72 & 0.23  \\
UGC 1281 & $5.5\pm0.3$, $>10$ & 0.73 & m22 & $4.4\pm1$ & 0.27 & m42 & 0.81  \\
AGC 122834 & $>10$ & 0.99 & m32 & $4.9\pm1$ & 0.01 & m72 & 2.78 \\
UGC 1561 & $\sim10$ & 0.41 & m42 & $7.9\pm1$ & 0.59 & m62 & 1.34  \\
NGC 855 & $>10$ & 0.45 & m72 & $7.5\pm1$ & 0.55 & m62 & 1.53 \\

\hline
\end{tabular}
%\end{minipage}
\end{footnotesize}
\end{center}
\end{table}

%{\bf For AGC 111945 and NGC 785, the metallicity of the 2nd burst is lower than that of the %1st!}

Below we discuss briefly individual objects. For each object we show the final B and net-H$\alpha$ images. Although we showed above that, at least from a kinematic point of view, the objects should be considered as parts of the same large-scale structure, in what follows we separate the discussion in the two ``classical'' galaxy groups recognized in the literature and in a few galaxies marked as ``isolated''.

\subsection {The NGC 672/IC 1727 group}
The NGC 672/IC 1727 group consists of six members: the NGC 672/ IC 1727 galaxy pair (VV338 pair) and four more nearby dIrr galaxies. Three of these (AGC  110482=KK13, AGC  111945=KK14 and AGC 111946=KK15) are included in the nearby dwarf galaxies' HI survey of Huchtmeier, Karachentsev \& Karachentseva (1997) and were previously associated with the NGC 672/IC 1727 group (e.g., Karachentseva \& Karachentsev 1998). The additional dwarf galaxy AGC  112521 was identified by the ALFALFA HI survey (Giovanelli et al.\ 2005) as a possible member of this group.   All the galaxies of this group, except for the NGC 672/IC 1727 pair, have very low HI masses: $log[M_{HI}]<7.5$ (Saintonge et al.\ 2008).

The NGC 672 and IC 1727 galaxies were both classified as late-type (e.g., Combes et al.\ 1980). NGC 672 was classified Sc$^{+}$ (Holmberg 1958) and later as SBc (de Vaucouleurs et al.\ 1976). IC 1727 was classified by the same authors as IrrI and SBm. It shows a disturbed spiral shape with a bright central ridge of blue condensations, possibly an incipient bar.

Saintonge et al.\ (2008) mentioned that the total HI mass of this group of galaxies is $2.5 \times10^{9}M_{\odot}$ of which 97$\%$ is concentrated in the NGC 672/IC 1727 pair. Previous HI analyses (Combes et al.\ 1980) using the OVRO two-element interferometer hinted that the galaxies were strongly interacting and that about 3$\%$ of the HI was contained in the connection between the galaxies. Note that this observation had an angular resolution of $\sim$2.5 arcmin. 34 HII regions were reported in NGC 672 and 28 in IC 1727 (e.g, Hodge 1982, Hodge \& Kennicutt 1983). %{\bf Noah: more on optical appearance of these two}

Both NGC 672 and IC 1727 have colours typical of field galaxies. These fitted the models reasonably well ($\chi^{2}< 3$), indicating two SF bursts at similar times for both galaxies. For both objects the first burst took place more than 10 Gyr ago, whereas the second burst took place only $\sim$7-8 Myr ago. In NGC 672, 57$\pm$10\% of the stars measured today were formed in the earlier burst. In IC 1727, 37$\pm$10\% of the stars were formed in the first (earlier) burst.

{\bf AGC 110482:} This object shows a fuzzy disky appearance with a brighter center and north-east side. The first star burst in this dIrr galaxy took place more than 10 Gyr ago and the second burst only $\thicksim$6 Myr ago, forming 70$\pm$10\% of the stars we observe today. Two HII regions can be identified in this galaxy, with line fluxes of 1.07$\times10^{-14}$~erg/cm$^{2}$/s and 7.82$\times10^{-15}$~erg/cm$^{2}$/s respectively; these produce $48\%$ of the total H$\alpha$ flux from this galaxy.

{\bf AGC 111945:} this dIrr galaxy shows up only as a low surface brightness blob with a slightly brighter central condensation. Since it does not show an obvious disk, it was not possible to measure the PA of its major axis. The galaxy has no prominent HII regions yet shows an H$\alpha$ line flux of $2.98 \times 10^{-15}$ erg/cm$^{2}$/s produced a diffuse H$\alpha$ component. The line emission is very low; this is why we do not show a nH$\alpha$ image in Figure~\ref{fig:AGC111945}. In addition, this object does not fit well the model predictions and has only $\chi^{2}$=9.17 for the best fit. The fit indicates that the first SF burst took place only 1.26$\pm0.3$ Gyr ago providing $50\pm10\%$ of the luminosity, whereas the second burst took place $\sim$4 Myr ago with  the younger population having a lower metallicity than the older one. The following nine fits imply a first burst $\sim$6.3 Gyr ago. In an attempt to improve the poor fit, models allowing more than two bursts were run; these indicated a first burst $8-10$ Gyr ago contributing only $\sim$10\% of the measured light, followed by several %{\bf (how many)}
 SF bursts with the most recent one taking place $\sim$4 Myr ago. Note that all alternatives require that an SF event take place a few Myr ago.

{\bf AGC 111946:} this galaxy shows a low surface brightness disk-like light distribution, elongated in an approximately North-South direction. It has F(H$\alpha$)=$3.33 \times 10^{-15}$ erg/cm$^{2}$/s. No specific HII regions were detected, though the nH$\alpha$ image shows a faint and elongated area of enhanced brightness. The region has F(H$\alpha$)=1.54$\times10^{-15}$~erg/cm$^{2}$/s, half the total line flux from this galaxy. The first SF burst took place $\sim$1.5 Gyr ago, contributing $50\pm10\%$ of the present luminosity. The following nine fits indicate another option with a first burst happening $\sim$10 Gyr ago. This is further supported by models allowing more than two SF bursts. The recent burst took place only $\sim$2.5 Myr ago.

{\bf AGC 112521:} this dIrr galaxy shows a very low surface brightness on the blue image. It has F(H$\alpha$)=4.39 $\times 10^{-15}$ erg/cm$^{2}$/s and shows a single star-like HII region south of the main diffuse body. The model fits indicate that the first SF burst took place more than 10 Gyr ago accounting for $37\pm10\%$ of the present luminosity, while the second burst took place $\sim$7 Myr ago. {\bf check pix}

Considering these galaxies as a single group, they show a variety of SFRs but the common property is that all objects experienced a very recent SF burst, in the last 10 Myrs. The pair NGC 672/IC 1727 has a much higher SFR ($\sim10^{-1}, 10^{-2}$ M$_{\odot}$/yr) than the other group  members. The galaxies of this pair are much more luminous, by 4-7 mag., than the other objects of this group. They are also more massive with w$_{50}(HI)\simeq$100-200 km sec$^{-1}$ec. The other galaxies, except AGC 110482 that has an SFR$\sim$10$^{-3}~M_{\odot}/yr$, have  SFR$\sim$10$^{-4}$ M$_{\odot}$/yr;  they are forming stars at a very low rate at the current epoch. %This is in accordance with the absence of HII regions in the latter galaxies. In addition, t
The B-V colours of those galaxies correspond to the typical range of field galaxies: $0\leq B-V\leq0.9$ (e.g, Heller, Almoznino \& Brosch 1999). All V-R and V-I colours, except AGC 112521, fall within $0.2\sim$V-R$\sim0.7$ and $0.2\sim$V-I$\sim1.0$. AGC 112521 has V-I$\sim$1.36 mag.

\subsection {The NGC 784 group}

Six galaxies are considered members of this group: NGC 784, AGC 111977=KK16, AGC 111164=KK17, AGC 122834, AGC 122835 and possibly UGC 1281. NGC 784, AGC 111977, AGC 111164 and UGC 1281 were previously associated into a single group (see Karachentsev et al.\ 2004, Tully et al. 2006). Saintonge et al.\ (2008) identified two additional HI clouds that could be group members: the dIrr galaxy AGC 122834  and AGC 122835, which has no known optical counterpart. The recession velocities of these HI objects are 51 and 30 km sec$^{-1}$ respectively, while the radial velocity of the other group members is 150-210 km sec$^{-1}$. %{\bf Where is the HI cloud? Plot on linear relation in figs 1 and 2 by using a different symbol?}

NGC 784 and UGC 1281 have been classified as Sdm, while UGC 1281 has at times not been considered a member of this group (e.g., Huchtmeier et al. 1997, Markova \& Karachentsev 1998, and Tikhonov \& Karachentsev 2006). Saintonge et al.\ (2008) mentioned that the smaller members of the group (all galaxies except NGC 784 and UGC 1281) have very low HI masses: $log[M_{HI}]<7.5$.

{\bf NGC 784:} the measured colours compared with the model predictions indicate that $31\pm10\%$ of the galaxy was created in a burst more than 10 Gyr ago, where the rest of the light is produced by stars formed $\sim$9 Myr ago. The colours do not fit well a two-burst model, with $\chi^{2}$=16.78 for the best fit caused mainly by an excessive V-R colour %{\bf (could this be because of improper line subtraction?}.
 The high H$\alpha$ flux (2.2 $\times 10^{-12}$ erg/cm$^{2}$/s) indicates a relatively high SFR ($1.95 \times 10^{-2} M_{\odot}/yr$). This galaxy also shows a large number of HII regions (at least $\sim20$). The nH$\alpha$ image (Figure~\ref{fig:NGC784}) shows a complex of HII regions just below the nucleus; a similar but fainter complex is visible in a diametrically opposite part of the galaxy (upper part in the figure). This might be caused by an almost edge-on circum-nuclear ring. In this galaxy, as found for AGC 111945, the fit shows that the more recent star burst had a lower metallicity than the older one.

{\bf AGC 111977:} the models indicate that 65$\pm$10\% of the light is produced by stars formed in a first SF burst $\sim$10 Gyr ago, and the rest in a relatively recent one $\sim$27 Myr ago. The following nine fits show a possible recent burst $\sim$1 Myr ago; this is further supported by models allowing more than two bursts. It seems that at least three SF bursts are needed in order to describe well the colours of this galaxy. No specific HII regions are identified, yet we measure F(H$\alpha$)=1.74 $\times 10^{-15}$ erg/cm$^{2}$/s from the object. The blue optical image is fan-shaped, wider in the North part that in the South, with the center appearing brighter than other locations on the galaxy image. The brightest part of the galaxy in the broad-band image is where the strongest line emission is detected.

{\bf AGC 111164:} this dIrr galaxy shows one central HII region with F(H$\alpha$)=6.72$\times10^{-15}$~erg/cm$^{2}$/s, which is $68\%$ of the total line flux measured from this galaxy. The model predictions indicate that the first SF burst took place more than 10 Gyr ago contributing $65\pm10\%$ of the present starlight, while the second burst took place $\sim$1 Myr ago. The blue image shows a low surface brightness disk with a definite brighter star-like center. No traces of spiral structure are seen, but it is possible that the disk has other, fainter, condensations.

{\bf UGC 1281:} this Sdm galaxy has at least 10 HII regions, which produce a significant $H\alpha$ flux. The models indicate a first SF burst $\sim$5.5 Gyr ago contributing $73\pm10\%$ of the luminosity, and a second burst $\sim$4 Myr ago. The following nine fits show a possible first burst more than 10 Gyr ago; this is supported by models allowing more than two bursts.

{\bf AGC 122834:} this dIrr galaxy has been measured by Saintonge et al. (2008) and, despite its very small recession velocity ($\sim$50 km sec$^{-1}$), it is potentially associated with the NGC 784 group at an average recession velocity of 180 km sec$^{-1}$. The optical image of this galaxy is very small ($\sim$15 arcsec) and shows an oval ring-like outer structure. The colours we measured are weird and the galaxy has a weak H$\alpha$ flux ($\sim$10$^{-16}$ erg/cm$^{2}$/s). The 10 model fits indicate a single SF burst more than 10 Gyr ago that accounts for at least $99\%$ of the measured light. A possible second burst could have taken place $\sim$5 Myr ago contributing only $\sim1\%$ of the observed light.

We consider that the HI cloud AGC 122835 is another member of this group given its celestial position and radial velocity. The cloud is very close to AGC 122834 but has no optical counterpart. Saintonge et al. (2008) calculated that its total HI mass very similar to that of AGC 122834 and showed that its HI velocity dispersion is $\sim$twice as large as that of AGC 122834.

This galaxy group also shows consistency in the recent SF burst times, which took place less than 30 Myr ago, with the most recent one in AGC 111164 and the earliest one 9 Myrs ago in NGC 784. AGC 122834 might be the exception and may not be a part of this group. %{\bf Noah: where is the HI cloud of Amelie? Does it have anything to do with the other guys?}

\subsection {Additional galaxies}

{\bf NGC 855}=HI021404.3+275302 is projected ~13' to the east of the NGC 784 group, about 5.5 Mpc from the NGC 784 group and 2.2 Mpc from the NGC 672 group (assuming only Hubble expansion). Karachentsev et al.\ (2004) associated it with the NGC 925 group with a similar radial velocity but located $\sim$9$^{\circ}\simeq$1.3 Mpc from NGC 855 itself. The large distances indicate that  NGC 855 is relatively isolated from the rest of the galaxies of the sample.

 NGC 855 has been classified as an elliptical galaxy with a polar ring (Wallington et al.\ 1988), but with (B-V)=0.7$\pm$0.1 mag it is bluer than expected for this morphological type. Its appearance on our B-band image is that of a disk with a brighter centre, which has an off-centre extension. Wallington et al. reported an HI ring around the galaxy detected in VLA observations and a complicated structure within the galaxy.  We could not resolve individual HII regions in the galaxy, yet it has a relatively high H$\alpha$ flux. The model comparison shows a first burst more than 10 Gyr ago that contributes 45$\pm$10\% of the luminosity and a second burst only $\sim$8 Myr ago that provides more than half the light.

{\bf UGC 1561} is located south of the two groups of galaxies, at $\delta\simeq$24$^{\circ}$. It is 2.4 Mpc away from the NGC 672 group and 5.7 Mpc from the NGC 784 group in 3D space and assuming only Hubble expansion.  It was classified as an isolated dIrr (Braun \& Burton 2001) and shows a strong H$\alpha$ flux. U1561 has four bright HII regions arranged at the corners of a parallelogram. Clockwise from the north-eastern HII region, the regions show, respectively, line fluxes of 1.81$\times$10$^{-14}$~erg/cm$^{2}$/s, 1.23$\times$10$^{-15}$~erg/cm$^{2}$/s, 1.04$\times$10$^{-14}$~erg/cm$^{2}$/s and 2.39$\times$10$^{-14}$~erg/cm$^{2}$/s, and contribute 25\% of the total H$\alpha$ flux measured from this galaxy. The blue image shows a number of other faint condensations within the disk. The models indicate a first SF burst more than 10 Gyr ago contributing $41\pm10\%$ of the luminosity and a second burst only $\sim$8 Myr ago.

{\bf LEDA 169957} was first measured in the ``Arecibo Slice'' survey (Spitzak \& Schneider 1998) and was later classified as a low mass dIrr (see Braun \& Burton 2001). It shows a disky light distribution with a brighter central condensation and a hint of faint spiral arm in its Northern side. Being four degrees to the south-west of the NGC 672/IC 1727 group and $\sim$150 km sec$^{-1}$ higher in velocity, implies that its distance from the NGC 672 group is $\sim$1.8 Mpc assuming only Hubble expansion. It has not been associated with this group in the past, and we assumed it is possibly isolated, but it does line up well with the linear relation shown in Figure~\ref{fig:gals_X_vs_Vel}. The colours indicate at least two SF bursts, a first one more than 10 Gyr ago contributing 35$\pm$10\% of the luminosity, and a second $\sim$20 Myr ago. Models allowing more than two bursts indicate an additional recent burst at $\sim$1 Myr. This galaxy might be a distant member of the NGC 672/IC 1727 group, despite its recent burst that was probably earlier than that of the other members of the group.

These three galaxies, apparently isolated, also show recent star formation, in the last $\sim$30 Myrs. This, therefore, is a general characteristic of all the galaxies in this part of the sky, apart from the confirmation that they all fulfill the primary requirement of being dwarf galaxies. The width of the 21-cm line listed in Table~\ref{gal} indicates that in most cases the galaxies indeed have low total masses. These characteristics will be used to understand the nature of this galaxy concentration.

\section {Discussion}
\label{sec.discuss}

The SFH of the galaxies was derived by comparing the measured galactic colours to GEMs, and the results are summarized in Table \ref{T2}. Most sample galaxies were apparently formed more than 10 Gyr ago, with some possibly forming somewhat later (6-10 Gyr ago). All galaxies, with the exception of AGC 122834, experienced a recent SF burst at similar epochs, $\sim$1 to a few tens of Myr ago. This could be interpreted as a result of interactions that were at least partially responsible for the SF bursts, or as the influence of an SF mechanism that might be at work synchronizing the SF in separate and non-interacting galaxies. AGC 122834 does not seem to have experienced a significant recent burst; this could imply that it is not a member of the NGC 784 group, or that it is sufficiently distant from the other galaxies not to be affected by whatever mechanism is synchronizing the SF in the other objects.

\begin{figure*}[ht]
\centering{
  \includegraphics[width=12cm]{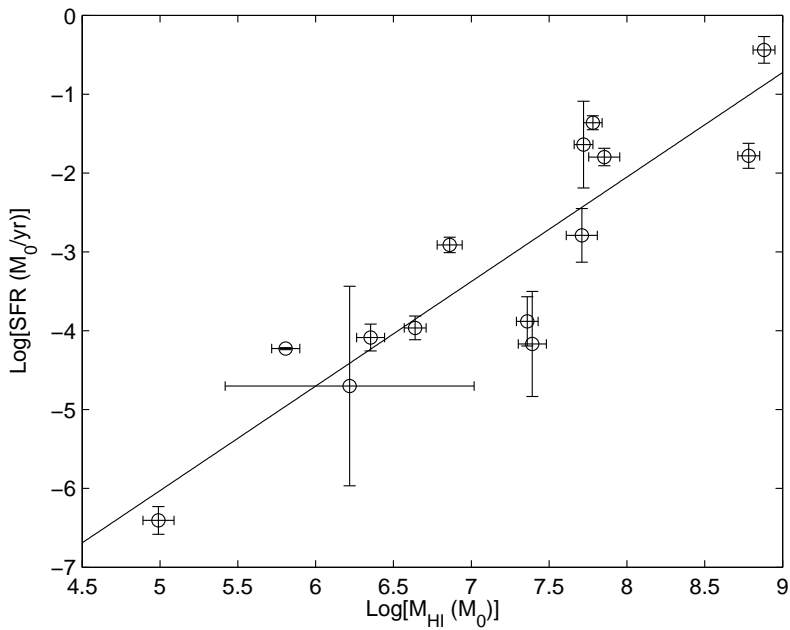}
  \caption{The relation between SFR and HI mass found for the objects studied here is similar to that measured in other environments for late-type galaxies.} \label{fig:SFRHI}}
\end{figure*}

\begin{figure*}[ht]
\centering{
  \includegraphics[width=12cm]{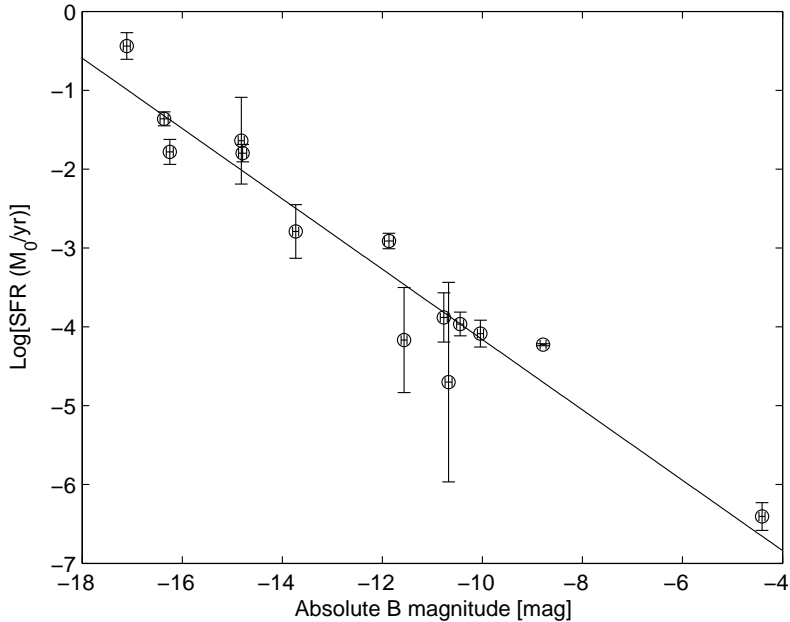}
  \caption{Relation between SFR and absolute B magnitude. A similar relation was found for galaxies in the nearby Universe (e.g., Karachentsev \& Kaisin 2007)} \label{fig:SFRVsmB}}
\end{figure*}

%\begin{figure}[ht]
%\centering{
%  \includegraphics[width=12cm]{SFRdensVsSFR.eps}
%  \caption{The relation between the SFR surface density and the SFR.} % % %\label{fig:SFRdensVsSFR}}
%\end{figure}

\begin{figure*}[ht]
\centering{
  \includegraphics[width=12cm]{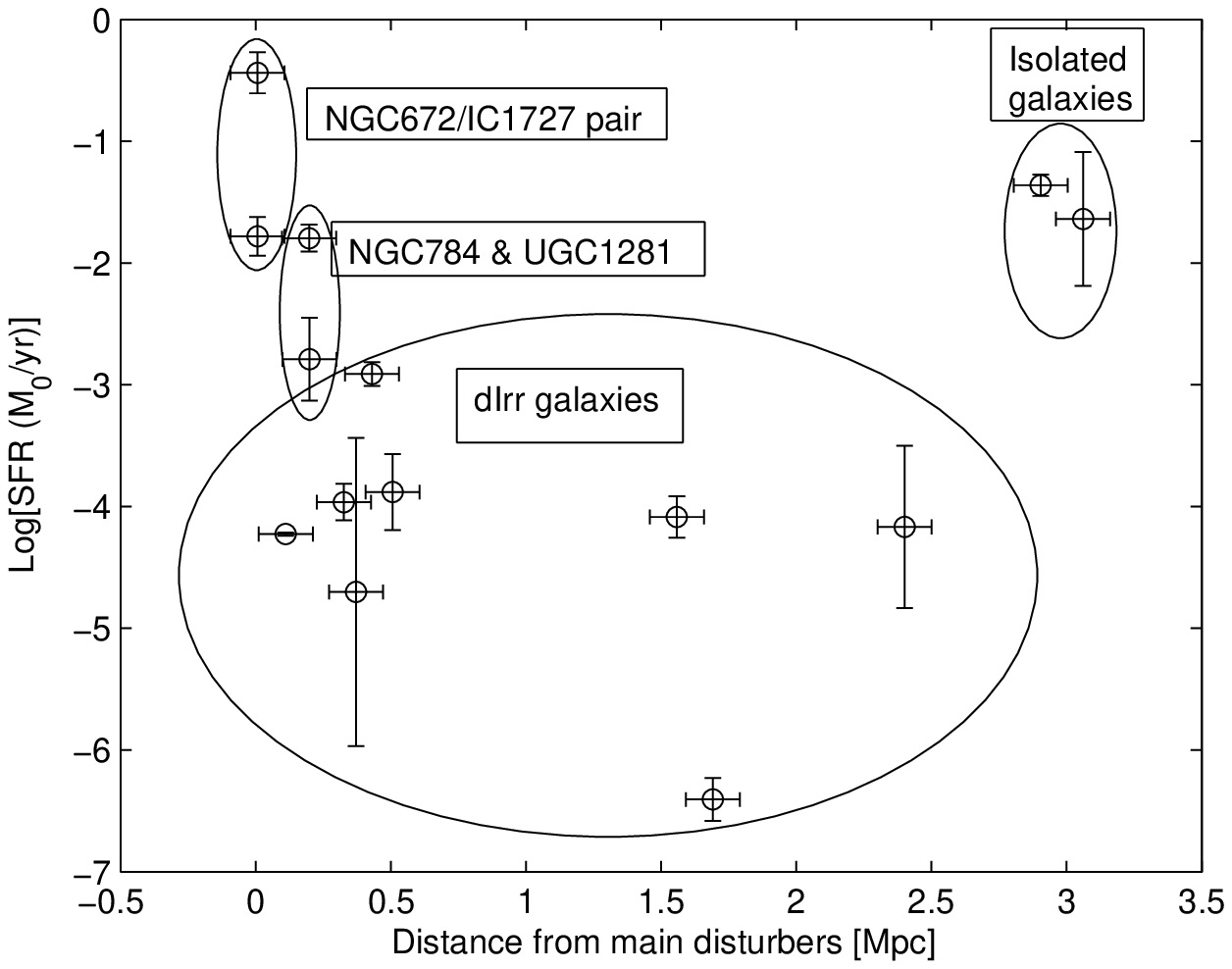}
  \caption{SFR vs. distance from the main galaxies. In this plot we consider the distances of each galaxy from the main galaxy in its group.} \label{fig:SFRvsDIST}}
\end{figure*}

Dwarf galaxies are common in the immediate vicinity of larger galaxies (Miller 1996, Cote et al.\ 1997). Some are believed to form as results of strong interactions or mergers between larger galaxies (e.g., Hunsberger et al.\ 1996, Bournaud \& Duc 2006) and can also be explained by hierarchical clustering scenarios that sometimes include supernovae winds (e.g., Dekel \& Silk 1986, Nagashami \& Yoshii 2004). Models can properly explain their observed colours only if the stars in these galaxies were formed in short bursts separated by long quiescence periods (e.g., Tosi et al.\ 1991 and references therein). The recent bursts could possibly be caused by the accretion of gas retained in the dark halos of these dwarfs (Dekel \& Silk 1986) or collected from intergalactic space (see below). %For most of the sample galaxies only two SF bursts %were required to explain their colours, where for a few, a better fit was obtained with models %allowing more than two bursts.

%\begin{figure*}[ht]
%\centering{
%  \includegraphics[width=12cm]{F_SFRVsD.eps}
%  \caption{SFR vs. distance from the main galaxies. In this plot we consider the distances of each galaxy from the main galaxy in its group.}
%\label{fig3Distr}}
%\end{figure*}

%\begin{figure}[ht]
%\centering{
 % \includegraphics[width=12cm]{F_SFRdensVsD.eps}
 %\caption{The behaviour of the SFR surface density with distance from main disturbers.} %\label{fig:SFRVsDens}}
%\end{figure}

We tested correlations of the individual SFRs against various parameters. Figure~\ref{fig:SFRHI} shows the relation between the SFR and the HI mass. The galaxies, though belonging to the same groups, show a variety of SFRs; this was found for other galaxy groups (e.g., Karachentsev \& Kaisin 2007). We found a linear correlation of $log[SFR]$ with $log[HI_{mass}]$, which implies that the more material is available for SF, the more intense the SF would be.  Kennicutt (1998), Taylor (2006) and others found that irregular and spiral galaxies follow the relation [SFR]$\propto M_{HI}^{1.4}$ (see also Karachentsev \& Kaisin 2007). The galaxies here, as Figure \ref{fig:SFRHI} shows, follow a relation with a similar slope of 1.3$\pm0.15$. %This small deviation in the slope can be explained by the small number of %galaxies in the sample.

We also found that $log[SFR]$ correlates with M$_B$, both derived assuming only Hubble expansion; see Figure~\ref{fig:SFRVsmB}). This indicates that the more intense the SF is, the more significant is the existing, relatively young, stellar population. Karachentsev \& Kaisin (2007) found a similar result for 150 galaxies in the local Universe where the relation was [SFR]$\propto$L$_{B}$. The galaxies in our sample follow a similar linear relation (Figure \ref{fig:SFRVsmB}) with a slope of $\sim$-0.4$\pm$0.04, which corresponds to [SFR]$\propto L_{B}^{1\pm0.1}$, similar to that found by Karachentsev \& Kaisin.

%The two correlations between $log[SFR]$ and absolute B magnitude or $log[M_{HI}]$ mass %indicate that the galaxies of the sample behave as normal galaxies and normal groups of %galaxies in the local Universe.

%{\bf Perhaps delete this paragraph that says nothing or do the comparison for a larger volume %defined by D$_{NN}$. Delete also Figures 4 and 6 that deal with areal SFR. } The average total %SFR density (i.e., per unit volume) at Z=0, has been found to be  0.02-0.03 %$M_{\odot}/yr/Mpc^{3}$ (see Martin et al.\ 2005, Hanish et al.\ 2006). Martin et al.\ (2005) %have found the luminosity function and SFR density in the near Universe by FUV and FIR data. %Hanish et al.\ (2006) have used H$\alpha$ data for over 300 galaxies to derive the local %Universe SFR density. The total SFR of the sample is $\thicksim$0.5 $M_{\odot}/yr$ in a volume %of 6 $Mpc^{3}$. This yields a SFR density of $\thicksim$0.08~$M_{\odot}/yr/Mpc^{3}$ which is %about 3-4 times higher than the averaged one reported by Martin et al.\ (2005) and Hanish et %al.\ (2006).

The SFR behaviour vs. the projected distance from the main galaxies of each group is shown in Figure~\ref{fig:SFRvsDIST}. The underlying assumption here is that the DGs might be formed in tidal interactions among the major galaxies in each group, thus the further away a DG would be from its ``main progenitor'' major galaxy, the weaker should its SF process be. It is difficult to evaluate the role of interactions in the triggering of SF in these groups, primarily since we do not have accurate 3D distances. The SFR is highest in the interacting pair NGC 672/IC 1727. The Sdm galaxies NGC 784 and UGC 1281, which might also be interacting, show relatively high SFR values as well. However, as Figure~\ref{fig:SFRvsDIST} shows, the six other DGs do not show a steadily decreasing SFR with (projected) distance from their ``main progenitors'', but rather a $\sim$uniform and low SF of $\sim10^{-4}$ M$_{\odot}$/yr. More isolated galaxies (that are also dIrr) show  independently high SFRs.

%{\bf This paragraph also says nothing and I suggest to delete it} Since the SFR depends on various local characteristics of a galaxy, more comprehensive conclusions could be made by examining the SFR surface density ($\sigma_{SFR}$) of each galaxy. The $\sigma_{SFR}$ was calculated for each galaxy by dividing the SFR by the projected area of each galaxy (at an isophote level of $\thicksim$25.5 mag/arcsec$^{2}$). We plotted $log[\sigma_{SFR}$] vs. $log[SFR]$ (Figure \ref{fig:SFRVsDens}); this shows a linear correlation. This shows that the SFR surface density***. In addition, it implies that $\sigma_{SFR}$ would behave in a similar way to the SFR with respect to the distance from the main disturbers. This is shown in Figure \ref{fig:SFRVsDens}. Again, the strongly interacting galaxy pair show the highest $\sigma_{SFR}$. The Sdm galaxies NGC 784 and UGC 1281, which are possibly interacting, show also relatively high $\sigma_{SFR}$, somewhat lower than that of the NGC 672/IC 1727 pair. The dIrr galaxies of each group show a decreased $\sigma_{SFR}$, whereas more distant galaxies (that can also be dIrr) show again independently higher $\sigma_{SFR}$.
%These more distant galaxies are isolated enough not to be affected and therefore the graph was re-plotted (Figure \ref{fig:SFRdensVs3D}) after "moving" these two galaxies to a zero distance, since each constitutes an individual body. This graph shows more clearly the decrease in the $\sigma_{SFR}$ of the dIrr galaxies with the distance from the main galaxies of each group, with $R^{2}\thicksim 0.6$.

We interpret this behaviour, along with the similar and recent SF burst times, as possible evidence for synchronicity in star formation in galaxies formed along a dark matter (DM) filament threading a nearby void. Brosch et al.\ (2004) found that galaxy interactions in dIrr galaxies are probably not a primary SF trigger and this seems to be the case here. The recent SF bursts in the dIrr galaxies of our sample, all taking place in the last few tens of Myrs, could be caused by accretion of extragalactic gas onto most of the galaxies. The accreted gas then forms stars upon collapse onto the disks. The lack of dependence of the SFR on the (projected)  distance from the brightest and most massive galaxies, and the timing similarity of the recent bursts, show that the potential main disturbers have minimal or nil influence on the dIrr galaxies in their vicinity. Another element, such as the postulated influx of extragalactic gas, could probably be the trigger of the synchronous SF.

%signature of probable interactions within the groups. These influences can be interpreted in a %number of ways: tidal interactions between the galaxies (accounting for their dark halos) %which caused the retained gas to collapse and form stars; past interactions which caused the %density field to be denser about the more massive galaxies and less dense in the outskirts of %each group, and with a consequence of more SF in the larger galaxies and less in the dIrr %around them. The latter might also be explained by past depletion of material towards the more %massive galaxies of the group.

At this point one remark is in order regarding our modeling the SF process as a collection of instantaneous bursts, separated by long quiescent intervals. The justification for this was that this SF mode is the one recognized to take place in dwarf galaxies, and all the objects here are dwarfs. However, if we examine the HI depletion time due to the star formation, by dividing the total HI mass of each object by its calculated SFR, we find that all objects could sustain SF at an almost constant level for long periods indeed, of order 10$^9$ yrs. Note that we have no information about the HI distribution; this could be very extended given the size of the ALFALFA beam, with only part of the HI taking part in the SF process. The long time needed to exhaust the HI reservoir could indicate that a quasi-continuous and low-level SF process might also explain the results. We emphasize, however, that the important issue here is that almost all objects show present-day SF.

The described behaviour and the proposed interpretation are, after all, not that surprising. It has been known that only strong interactions at short separations between galaxies of comparable masses enhance SF, and that the SFR increases as the strongly interacting galaxies come closer (e.g., Icke 1985). This could perhaps be the case in the NGC 672/IC 1727 (VV338) pair, but not in most of the galaxies studied here. In addition, it has been known that fairly isolated galaxies (in relatively denser environments) show generally higher SFR the more isolated they are. This could perhaps be the case for NGC 855 and UGC 1561. %Moreover, it was found that galaxies in denser %environments such as in galaxy groups show relatively low SF, as the dIrr galaxies of the %groups show. Our results are compatible with all these conclusions.

Kere\v{s} et al. (2005) showed, using hydrodynamic simulations, that low-mass galaxies with baryonic mass M$_{gal}\leq10^{10.3}$M$_{\odot}$ (or halo mass M$_{halo}\leq 10^{11.4}$M$_{\odot}$) can accrete intergalactic gas in a ``cold mode'', with the accreted gas colder than 10$^5$K thus capable of directly forming stars. The cold accretion in the simulations is often directed along DM filaments, allowing galaxies to efficiently draw gas from large distances. Such DM filaments form in N-body simulations such as those presented by Hahn et al. (2007), where it was also shown that that haloes in filaments are more oblate than cluster halos at high masses and that haloes in filaments tend to have their spin vectors pointing along the filaments.

The prediction regarding the spin alignment result of Hahn et al. (2007) is that the major axes of disky galaxies in the filament discussed here should be approximately perpendicular to the line along which the galaxies appear to be arranged. We checked this by estimating the position angles of the major axes of the galaxies (listed in Table~\ref{gal}) and comparing them to the general direction of the projected filament. The distribution of PA values shows two preferred values: one, with seven galaxies is centered at PA$\simeq$305$^{\circ}\pm12^{\circ}$. The other peak is broader and is centered at PA$\simeq$34$^{\circ}\pm31^{\circ}$. The general direction of the galaxy filament, using the same convention of measuring PAs, is $\sim$303$^{\circ}$. The conclusion is therefore that the Hahn et al. (2007) prediction regarding the correlated galaxy spins is fulfilled in a mixed-up way for this nearby galaxy alignment, with seven objects having their spin approximately aligned with the general direction of the filament and the other objects with their spin axes approximately perpendicular to the filament direction.

Dekel \& Birnboim (2006) showed that low-mass galaxies are built by cold gas streams in haloes below a critical shock-heating mass M$_{shock}\leq10^{12}$M$_{\odot}$, not by intergalactic gas heated by a virial shock. This accretion mode yields efficient star formation in low-mass galaxies. The only ingredient then needed to understand the galaxies studied here is to assume that the phenomenon can take place not only at z$\geq$2, but also at z$\approx$0. We could, therefore, assume that most of the 14 galaxies studied here do line up along a DM filament in a nearby void, and that they all exhibit a present-day synchronous star formation burst triggered by cold gas accretion from intergalactic space that is being focused by the DM filament.

\section{Summary}
\label{sec.summary}

We presented U, B, V, R, I, H$\alpha$ and NUV photometric measurements of 14 fairly isolated  galaxies in the local Universe. The galaxies are in the same sky area, most are arranged on a linear configuration, and have radial velocities $\leq$610 km sec$^{-1}$. 11 of these galaxies, and one HI cloud with no optical counterpart, were previously assigned to two groups of galaxies. All visible galaxies are ``dwarfs'' with M$_B\geq$--18 and can be morphologically classified as ``very late type'', including the dominating objects in each group. Three other galaxies were often considered to be field galaxies but are similar in absolute magnitude and classification to the other objects. The objects appear to form a single kinematically well-behaved ensemble that does not separate naturally into two groups. We derived the SFR and SFH of each object using photometry in the various bands and examined these properties in the context of the galaxy environment.

The galaxies show relations between the SFR and the HI mass, or m$_B$, known for field galaxies. The SFHs imply that similar low-intensity SF bursts took place in most galaxies a few Myr ago, despite their being spread along a $\geq$0.5 Mpc long feature. The behaviour of the SFRs, which do not decline steadily with increasing distance from the brightest galaxies in each group,  indicates that interactions do not affect the SF in these galaxies to the extent one would expect from strongly interacting galaxies. In particular, no signs of strong gravitational interactions, such as tidal tails or global disturbances, are observed. The main galaxies do appear to be interacting but on a one-to-one basis and do indeed show enhanced SF. Other dIrr galaxies in the same vicinity show only low SFRs. More isolated galaxies show independently high SFRs.

We propose that the observational evidence argues in favor of interpreting the galaxies as located on a DM filament that is itself located in a low-galaxy-density region, and is accreting intergalactic cold gas focused by the filament. We are therefore witnessing in the classical NGC 672 and NGC 784 groups of relatively bright and nearby late-type galaxies the basic phenomenon of hierarchical clustering, the direct formation and growth of small galaxies out of intergalactic gas accreted on a dark matter ``backbone''. This nearby galaxy collection offers, therefore, an ideal opportunity to study the phenomenon of hierarchical clustering in significant detail.

One possible conclusion from our proposed interpretation is that the gas being accreted could be observable in the vicinity of the galaxies; a way to detect it would be through very low column density detection of Lyman $\alpha$ absorption lines at the redshifts of the galaxies in spectra of background QSOs. Another would be to search for diffuse HI to much lower limits than achieved during the ALFALFA survey; it is unlikely that an extension of the ALFALFA scans will bring useful returns and perhaps this is a project that should wait for the entrance in operation of SKA.

\section{Acknowledgements}
This work is based on HI measurements collected at the Arecibo Observatory for the ALFALFA survey. The Arecibo Observatory is part of the National Astronomy and Ionosphere Center which is operated by Cornell University under a cooperative agreement with the National Science Foundation. We are grateful for access to the ALFALFA data and acknowledge the leadership of the ALFALFA survey's Principal Investigators Martha Haynes and Riccardo Giovanelli, and the valuable contributions of the other ALFALFA team members. This research used the facilities of the Canadian Astronomy Data Centre operated by the National Research Council of Canada with the support of the Canadian Space Agency. Amelie Saintonge is acknowledged for providing the data of the fall part of the sky where the galaxies studied here are located. Helpful discussions with Stefan Gottl\"{o}ber and Yehuda Hoffman are acknowledged. We also acknowledge the usage of the HyperLeda database (http://leda.univ-lyon1.fr).

\end{document}